\documentclass[acmsmall]{acmart}
\AtBeginDocument{%
  \providecommand\BibTeX{{%
    \normalfont B\kern-0.5em{\scshape i\kern-0.25em b}\kern-0.8em\TeX}}}

\setcopyright{acmcopyright}
\copyrightyear{2018}
\acmYear{2018}
\acmDOI{10.1145/1122445.1122456}

\acmJournal{JACM}
\acmVolume{37}
\acmNumber{4}
\acmArticle{111}
\acmMonth{8}

\usepackage{multirow}
\begin{document}

\title{Modular Tree Network for Source Code Representation Learning}

\author{Wenhan Wang}
\affiliation{%
  \institution{Peking University}
  \country{China}
}
\email{wwhjacob@pku.edu.cn}

\author{Ge Li}
\affiliation{%
  \institution{Peking University}
  \country{China}
}
\email{lige@pku.edu.cn}
\authornotemark[1]

\author{Sijie Shen}
\affiliation{%
  \institution{Peking University}
  \country{China}
}
\email{sjshen@pku.edu.cn}

\author{Xin Xia}
\affiliation{%
 \institution{Monash University}
 \country{Australia}}
\email{xin.xia@monash.edu}

\author{Zhi Jin}
\affiliation{%
  \institution{Peking University}
  \country{China}
  }
\email{zhijin@pku.edu.cn}
\authornote{corresponding author}

\renewcommand{\shortauthors}{Wang et al.}
\begin{abstract}
Learning representation for source code is a foundation of many program analysis tasks. In recent years, neural networks have already shown success in this area, but most existing models did not make full use of the unique structural information of programs. Although abstract syntax tree-based neural models can handle the tree structure in the source code, they cannot capture the richness of different types of substructure in programs. In this paper, we propose a modular tree network (MTN) which dynamically composes different neural network units into tree structures based on the input abstract syntax tree. Different from previous tree-structural neural network models, MTN can capture the semantic differences between types of AST substructures. We evaluate our model on two tasks: program classification and code clone detection. Our model achieves the best performance compared with state-of-the-art approaches in both tasks, showing the advantage of leveraging more elaborate structure information of the source code.
\end{abstract}



\maketitle

\section{Introduction}
Sour code representation learning is a set of techniques that allow a system to automatically discover the representations needed for feature detection or classification from source code. Due to the significant progress in deep learning in recent years, modeling source code by deep neural networks has drawn more and more attention. Deep-learning based source code representation techniques have achieved considerable success in many tasks in program analysis, including program classification \cite{mou2016convolutional,liang2018comment}, defect prediction \cite{yang2015deep,wang2016automatically}, code clone detection \cite{white2016deep,wei2017supervised}, code summarization \cite{iyer2016summarizing,allamanis2016convolutional,liang2018comment}, and malware detection \cite{yuan2014droid}.

As Hindle et al. \cite{hindle2012naturalness} has shown, programming languages have many similar statistical properties with natural language. So when it comes to representing programs using deep learning models, researchers tend to use models that have already shown success in natural language processing rather than building new models for the source code. Some basic approaches \cite{bhoopchand2016learning,li2017code} split the source code into a sequence of tokens like natural language sentences, where their models include convolutional neural network (CNN) \cite{allamanis2016convolutional,huo2016learning} and recurrent neural network (RNN) \cite{iyer2016summarizing,bhoopchand2016learning,li2017code}. However, different from natural languages, whose structure is relatively simple, a programming language contains complex and explicit structural information \cite{mou2016convolutional,allamanis2018survey}. These pieces of structural information can be extracted from a program's abstract syntax tree (AST). An example of an AST in C programming language and its corresponding code snippet is presented in Figure 1. An AST consists of a group of semantic units, each one include a non-leaf AST node and all its children. Different types of semantic units have different semantic meaning. For example, In Fig.1, {\fontfamily{\ttdefault}\selectfont While} nodes have two children: a condition node and a loop body node, together they build a loop sketch. The children of {\fontfamily{\ttdefault}\selectfont BinaryOp} nodes are two operands, and together they form a calculation on identifiers. Each type of semantic unit is defined by a grammar production rule. As the grammar of a programming language contains a finite set of production rules, the total number of semantic unit types in a programming language is also finite. Abstract syntax trees are constructed from different semantic units, which form the complex structure of the source code.

\begin{figure}[htb] 
\centering 
\includegraphics[height=3.5cm, width=8.5cm]{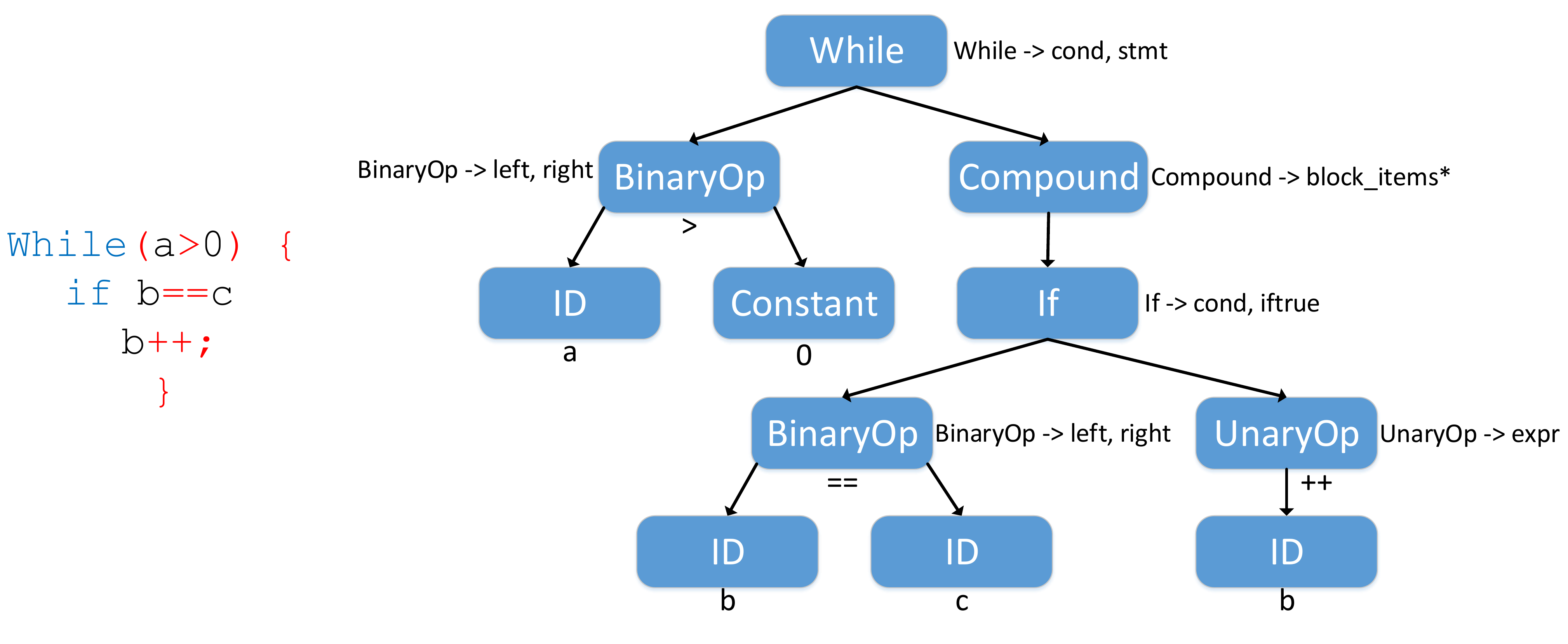} 
\caption{A simple C code snippet and its corresponding abstract syntax tree fragment. Beside each non-leaf node shows the production rule related to the semantic unit which contains this node and its children. Some nodes, mainly identifiers, contain one or more values, while other nodes only contain a node type.}
\end{figure}

There have already been several tree-structured neural network models for representing tree-structured data, including Tree-LSTM \cite{tai2015improved} and TBCNN \cite{mou2016convolutional}. Although these models utilize the tree structure of input data, they share a common weakness: they use the same neural network unit to model different kinds of AST semantic units. Although this feature makes these models available for trees from arbitrary data domains, it also prevents them from handling the differences between semantic units. Using only one neural network unit to capture the meaning of all semantic units is apparently not enough, which can cause underfitting.

In this paper, we propose Modular Tree Network (MTN), a dynamically-composed tree-structured neural network framework which models different AST semantic units with different neural network modules. We select a group of semantic units related to program structure based on the grammar of programming languages and design a unique neural module for each of them. For an input AST, the network architecture is constructed with these neural modules based on the specific structure of the AST. Then the vector representation of the input AST is calculated in a bottom-up manner.

We evaluate the performance of our model on two software engineering tasks, i.e., program classification, and clone detection. In program classification, we propose a 293-classes POJ dataset and evaluate our model on it. Results show that our model achieves 1.8\% higher accuracy than the state-of-the-art approaches. In clone detection, we compare our model with CDLH on OJClone dataset, and our model improves the F1-score from 0.57 to 0.90.

The main contributions of this paper are as follows:

(1) We propose a tree-structured neural network framework composed of multiple submodules for learning a representation of source code fragment based on its AST. 

(2) We employ this model and the representation it learns to complete two tasks on source code analyzing: program classification and code clone detection. 

(3) Our proposed model achieves better results compared to previous neural network models, showing that utilizing more detailed semantic information of source code is advantageous. We also show that MTN converges faster than baseline models and can learn better on small datasets.

{\bf Paper Organization. } The remainder of the paper is organized as follows. We describe the background of deep learning in Section 2. We present the structure of our model MTN in Section 3. We describe the application of MTN on two program analysis tasks in Section 4. We demonstrate and analyze our experiments and results in Section 5. We show the threats to validity in Section 6 and list related works in Section 7. Finally, in Section 8 we make a conclusion on our work.

\section{Background}
In this section, we briefly introduce the background of our model MTN: tree-structured neural networks, especially recursive neural network and tree-LSTM.

\subsection{Recursive Neural Network (RvNN)}
Recursive neural network (RvNN) is the most basic type of tree-structured neural networks. Suppose a node $j$ in a binary tree has two children $c1$ and $c2$, its representation is computed from its children by:
\begin{equation}
h_{j}=tanh(W[h_{c1}:h_{c2}]+b)
\end{equation}
$W$ is a weight matrix, and $b$ is a bias term. Since $W$ is shared across all non-leaf nodes in a tree, recursive neural networks require trees with a definite branching factor, e.g., binary trees. Besides, in recursive neural networks, only leaf nodes take input vectors, so it is mostly used in structures like natural language constituency trees, where the lexical tokens of a natural language sentence are all stored in their leaf nodes.
\subsection{Tree-LSTM}
Tree LSTM is a generalization of Long short-term memory (LSTM) to tree-structured network topologies \cite{tai2015improved}. LSTM is a variant of recurrent neural networks (RNNs) proposed to solve the problem of long-term dependencies existed in traditional RNNs. In time step $t$, a traditional RNN cell calculates the hidden state of the current time step $h_{t}$ by the input of current time step $x_{t}$ and the hidden state of the previous time step $h_{t-1}$, using the below equation:
\begin{equation}
h_{t}= tanh(Wx_{j}+Uh_{t-1}+b)
\end{equation}
where $W$ and $U$ are different weight matrices for input vector and hidden state. An LSTM unit extends traditional RNN cell with a forget gate $f$, an input gate $i$, an output gate $o$, and a memory cell. The forget gate decides what information of the previous memory cell we choose to forget, the input gate controls the extent of using new information to update the memory cell, and the output gate controls the extent to which the value in the cell is used to compute the hidden state of the LSTM unit. Each gate takes $x_{j}$ and $h_{t-1}$ with a different set of parameters $W$, $U$ and $b$. Tree-LSTM shares these gates and cell with traditional LSTM but for a node $j$ in a tree, instead of taking the hidden state of the previous time step, a tree-LSTM unit takes the hidden states of all of its children and feed a combination of these states into its input and output gate. For the forget gates, a tree-LSTM unit contains multiple forget gates with shared parameters, and each one only receives the hidden state of its associated child. 

Under these settings, there exist two versions of tree-LSTM: Child-Sum Tree-LSTM and N-ary Tree-LSTM. We will briefly introduce these two variants and analyze their drawbacks.
\subsubsection{Child-Sum Tree-LSTM}
In child-sum tree-LSTM, for a node $j$, we sum up the hidden states of all its children and use this summation as the input hidden state of the input gate and output gate. Child-sum Tree-LSTM can be built on arbitrary tree structures. However, since changing the order of children do not affect the result of the summation operation, its calculation process loses the information of the order of children.
\subsubsection{N-ary Tree-LSTM}
Similar to recursive neural networks, n-ary Tree-LSTM can only be applied on tree structures where the branching factor is at most N. Its main difference from child-sum tree-LSTM is that n-ary tree-LSTM aims to leverage the order of children. Instead of summing up all children hidden states like child-sum tree-LSTM, n-ary tree-LSTM associate each child with a different weight matrix $U$. By introducing different parameters for different children, n-ary tree-LSTM can capture more fine-grained semantic information, such as the difference in importance between children of different positions.

When applying n-ary tree-LSTM to trees with indefinite branching factors, like abstract syntax trees, one need to convert them into n-ary trees (often binary). This process not only destroys the original syntax relationships between nodes in the original AST but also make ASTs deeper, which may cause long-term dependency problems.

\section{Approach}
Our proposed model, modular tree network (MTN), refines the basic structure of child-sum tree-LSTM by replacing a part of tree-LSTM units with neural modules designed for different AST semantic units. This idea of building different neural modules for different semantic units can be applied to any programming languages. However, we need to design different neural modules for semantic units in different programming languages, because:

1. Different programming languages may contain different semantic units. E.g., C contains {\fontfamily{\ttdefault}\selectfont Case} statements while Python does not have them, Java has {\fontfamily{\ttdefault}\selectfont try ... catch} statements while C does not have them.

2. Similar semantic units in different programming languages may have different form. For example, in the AST of C, a {\fontfamily{\ttdefault}\selectfont For} statement is defined by the production rule $For \mapsto (init, cond, next, stmt)$, while in Python the production rule for the {\fontfamily{\ttdefault}\selectfont For} statement is $For \mapsto (target, iter, body*, orelse*)$ (which "orelse" seldom appears).

In this paper, we only discuss MTN for C as an example, but the proposed model can be generalized to other programming languages by designing new neural modules for different semantic units. The overall framework of MTN is shown in Figure 2. The input of MTN is the AST representation of source code, and the output of MTN is a vector representation of the AST. In the following, we show the general structure of MTN units and define a set of neural modules for semantic feature capturing in C programming language. Then, we show the whole process of applying MTN model to calculate vector representations for ASTs.
\begin{figure*}[tbp] 
\centering 
\includegraphics[height=5cm, width=15cm]{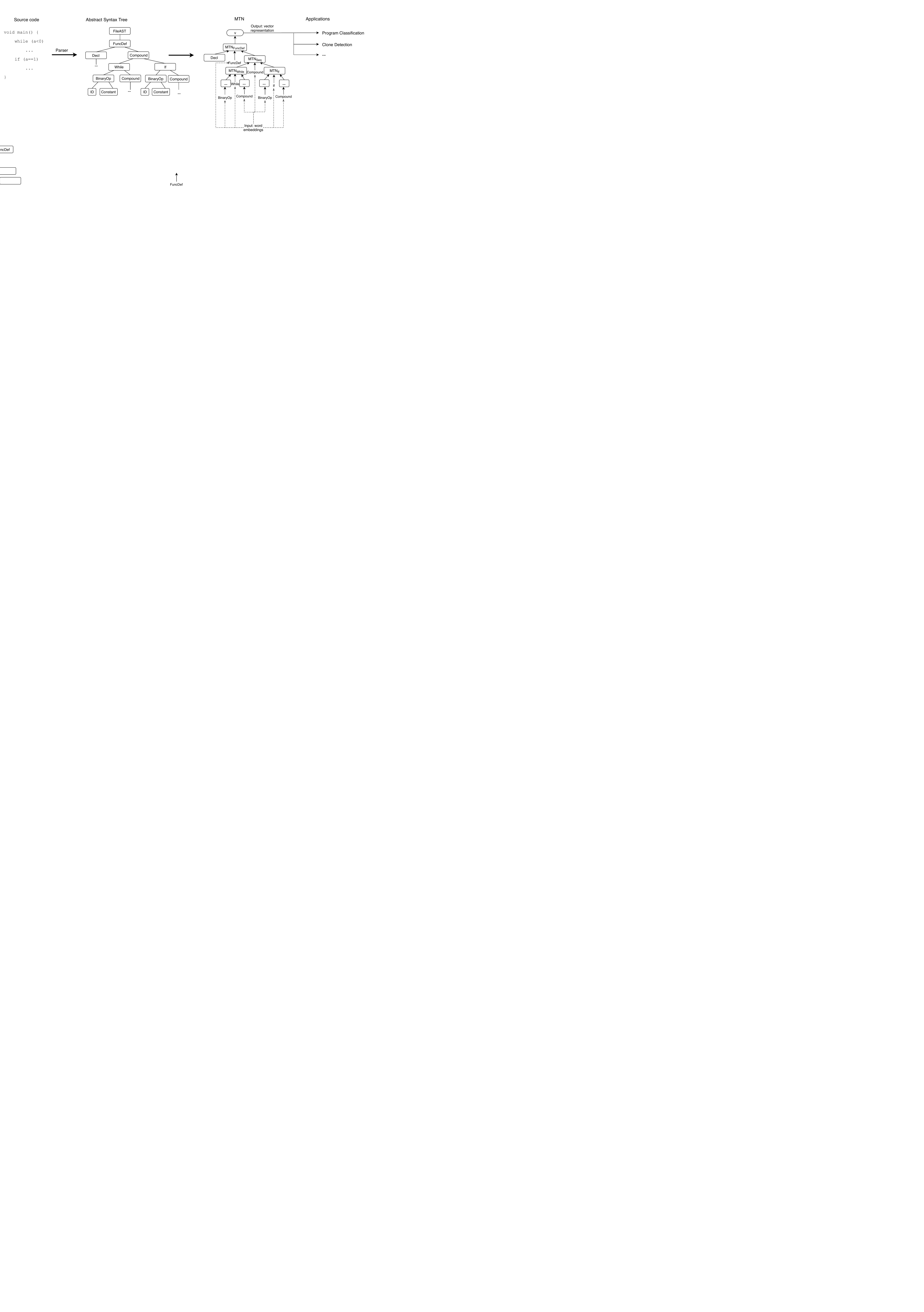}
\caption{The overall process of calculating vector representation for source code using MTN.} 
\end{figure*}
\subsection{General Structure of MTN Unit}
In MTN, we replace the operation of summing up all children hidden states in Child-Sum Tree-LSTM by applying one of our designed neural modules. Given a node $j$, an MTN unit calculates its hidden state by:

\begin{align}
& \widetilde{h}_{j}=F_{type-j}(h_{s_{1}},...,h_{s_{n_{j}}}) \\
& i_{j}=\sigma (W^{(i)}_{type-j}x_{j}+U^{(i)}_{type-j}\widetilde{h}_{j}+b^{(i)}_{type-j})  \\
& f_{jk}=\sigma (W^{(f)}_{type-j}x_{j}+U^{(f)}_{type-j}h_{k}+b^{(f)}_{type-j})  \\
& o_{j}=\sigma (W^{(o)}_{type-j}x_{j}+U^{(o)}_{type-j}\widetilde{h}_{j}+b^{(o)}_{type-j})  \\
& u_{j}=\textup{tanh}(W^{(u)}_{type-j}x_{j}+U^{(u)}_{type-j}\widetilde{h}_{j}+b^{(u)}_{type-j})  \\
& c_{j}=i_{j}\odot u_{j}+\sum_{k\in C(j)}f_{jk}\odot c_{k}  \\
& h_{j}=o_{j}\odot \textup{tanh}(c_{j}) 
\end{align}

Here, $W$, $U$, $b$ are parameters associated with the type of $j$ and $\odot $ is the element-wise product operator. In equation (3), MTN passes these hidden states through a neural network module designed by us. $F_{type-j}$ is a unique neural module for the node type of $j$ which receives the hidden states for all children of $j$ and produce one vector as an intermediate result. We will describe the details of $F_{type-j}$ in section 3.2. For the sake of time efficiency and implementational complexity, we only design modules for several kinds of semantic units. For the rest kinds of the node without specially designed neural modules, we add up their children's hidden state like Child-Sum Tree-LSTM instead of applying a module $F$. Equations (4) to (9) are the same as child-sum tree-LSTM, which act as the input gate, forget gates, output gate and the memory cell for the MTN unit.

An MTN unit can be seen as a neural module (equation (3)) wrapped by a Tree-LSTM-like "container" (equations (4)-(9)) which contains a set of gates and memory cells. With MTN hidden size $d$, a "container" has $8d^{2}+4d$ parameters, which is the same as child-sum tree-LSTM. For the parameters of gates and memory cells in the "containers," we propose two strategies of parameter sharing:

\textbf{MTN-a}: $W$, $U$, $b$ (the container parameters) in equations (4)-(7) are shared across all MTN units in a program AST.

\textbf{MTN-b}: We take the idea of using different network architecture and parameters from neural modules to the whole MTN unit.For each type of semantic unit that has a unique $F_{type}$ module, we associate it with a different set of corresponding container parameters $W$ $U$ and $b$ along with different $F_{type}$. The set of parameters of a particular type of semantic unit are shared among all occurrences of this semantic unit type. For those subtrees without $F_{type}$, they share another set of container parameters. 

An intuitive demonstration between MTN-a and MTN-b is shown in Figure 3. In this figure, different colored boxes indicate different container parameters.

\begin{figure}[htbp] 
\centering 
\includegraphics[height=4.5cm, width=11cm]{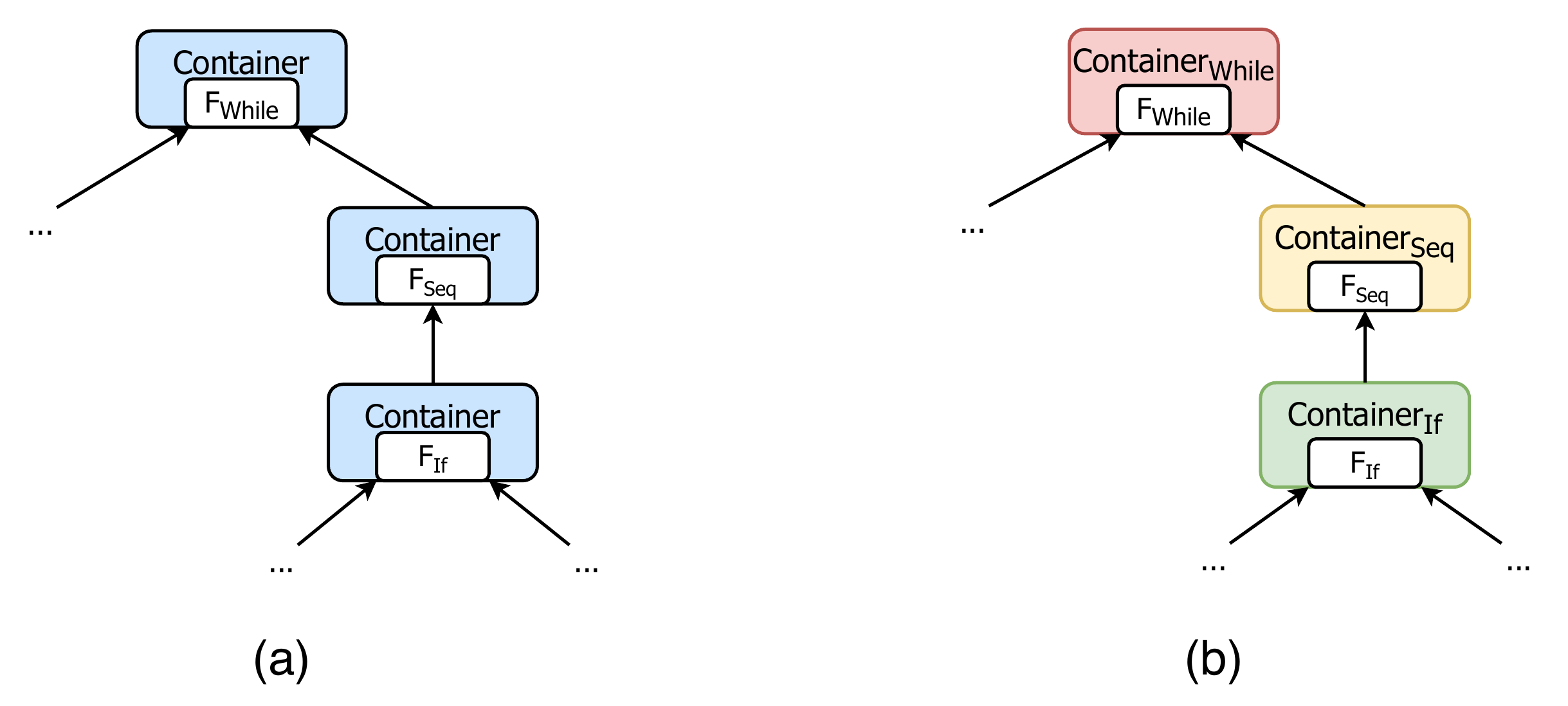} 
\vspace{-1em}
\caption{The two parameter sharing strategies of MTN. (a): MTN-a. (b): MTN-b}
\vspace{-1em}
\end{figure}

\subsection{Neural Modules For Different Semantic Units}
The semantic units of C can be divided into several categories: 
\begin{itemize}
\item Units which define logic structure and control flow information, like branches and loops.
\item Units contain data information, like a variable declaration.
\item Units contain calculation information, like operators and variable assignments in expressions.
\end{itemize}
In this paper, we mainly focus on semantic units which represent the basic structure of programs. We design 8 different neural network modules with different parameters for different semantic units. First there are 7 kinds of node related to program logic structure: {\fontfamily{\ttdefault}\selectfont FuncDef}, {\fontfamily{\ttdefault}\selectfont While}, {\fontfamily{\ttdefault}\selectfont DoWhile}, {\fontfamily{\ttdefault}\selectfont For}, {\fontfamily{\ttdefault}\selectfont If}, {\fontfamily{\ttdefault}\selectfont Switch} and {\fontfamily{\ttdefault}\selectfont Case}. Apart from these kinds of nodes we mentioned before, there are also a group of nodes which the number of their children is not definite, but their children are a sequence of nodes. We also design a neural module $F_{seq}$ for these nodes. We show the whole set of modules in Table 1 and report the number of parameters of these modules given the hidden size $d$. In the following part, we will give the implementation of these neural modules and shed some light on the thoughts behind the structure of them.
\subsubsection{FuncDef, While, DoWhile and Switch}
These four types of nodes are similar in these ways: they all have two children, and the semantics of their left children differs from their right children. So we use a recursive neural network module to model them. In order to address the semantic differences between these four types of nodes, we set different parameters $W$ and $b$ for their modules.
\begin{figure}[h] 
\centering 
\includegraphics[height=2cm, width=6cm]{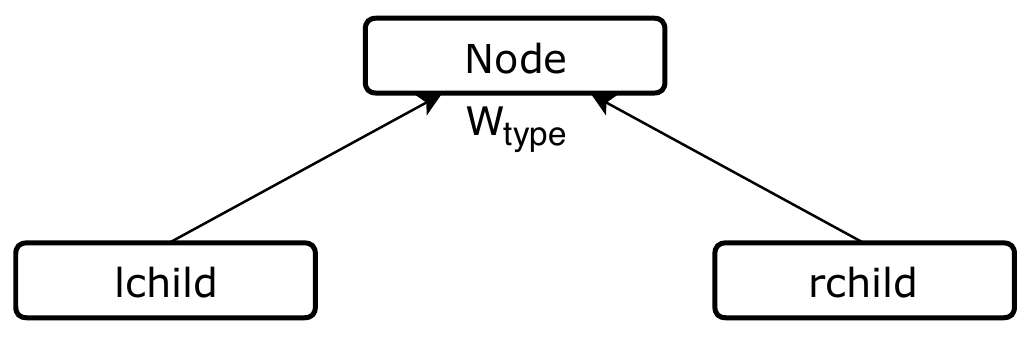} 
\caption{Structure of FuncDef, While, DoWhile and Switch modules. } 
\end{figure}

\subsubsection{For}
A {\fontfamily{\ttdefault}\selectfont For} node has four children: an initialization statement, a condition statement, a next iteration statement and a loop body. The first three parts form the controller of a loop, so we use an RvNN unit to model them, and feed the intermediate output of this unit and the state of the loop body node to another RvNN unit. Figure 4 shows the structure of our For module.
\begin{figure}[h] 
\centering 
\includegraphics[height=2.8cm, width=6cm]{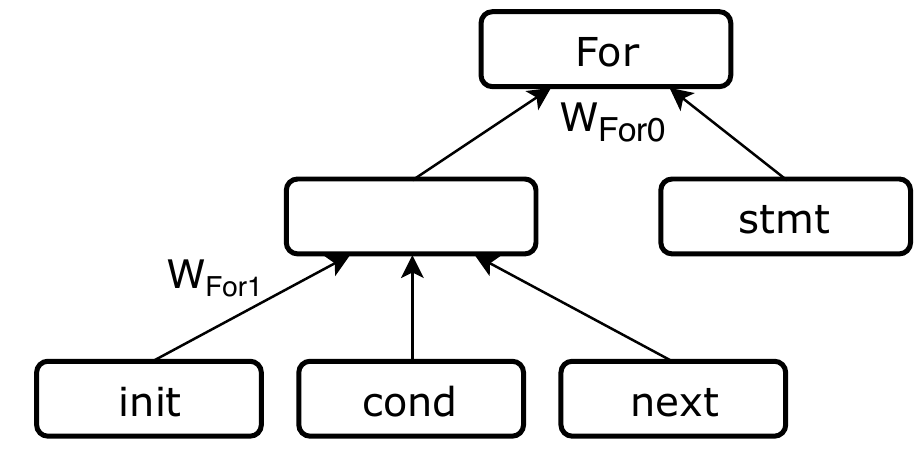} 
\caption{Structure of For module. } 
\end{figure} 
\subsubsection{If}
Different from previously mentioned AST nodes in C, an {\fontfamily{\ttdefault}\selectfont If} node does not always have the same number of Children. An {\fontfamily{\ttdefault}\selectfont If} node can have 2 or 3 children, depending on whether it contains an else statement. If an {\fontfamily{\ttdefault}\selectfont If} node does not contain an else statement, we simply use a recursive neural network to combine its condition node and branch body node. If an {\fontfamily{\ttdefault}\selectfont If} node has an else statement, we separately combine the condition node with the iftrue statement and the iffalse statement with two recursive units with the same parameters. Then we compute the element-wise max for the output of the two recursive units to get the final output of the {\fontfamily{\ttdefault}\selectfont If} module.
\begin{figure}[h] 
\centering 
\includegraphics[height=2.8cm, width=8cm]{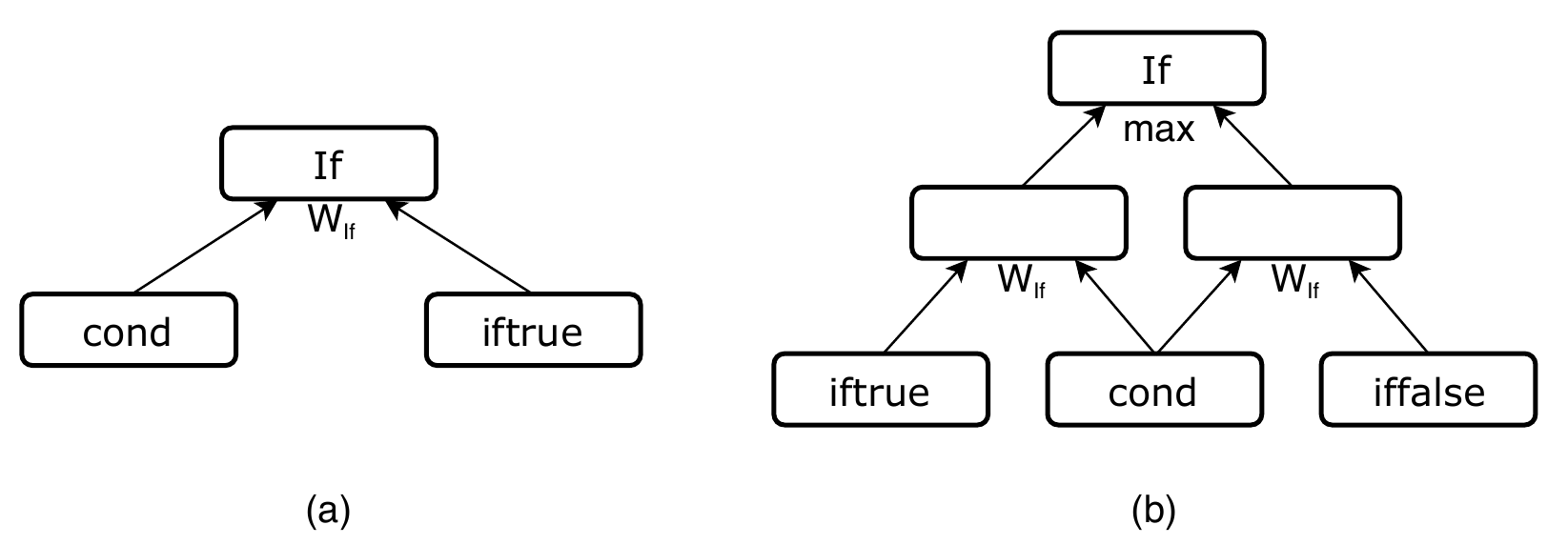} 
\caption{Structure of If module. (a): without an else statement. (b): with an else statement.} 
\end{figure}
\subsubsection{Case}
Since Case nodes are a component of switch-case statements, we also design a module for them. The children of a Case node can be divided into two parts: a constant expression which is the label of the current Case branch, and the body of the branch which is a sequence of statements. Since the statements of a Case branch is executed sequentially, we use an LSTM network to encode the root nodes of these statements. The last hidden state of this LSTM and the state of the constant expression node are passed into a recursive NN layer.
\begin{figure}[h] 
\centering 
\includegraphics[height=4cm, width=6cm]{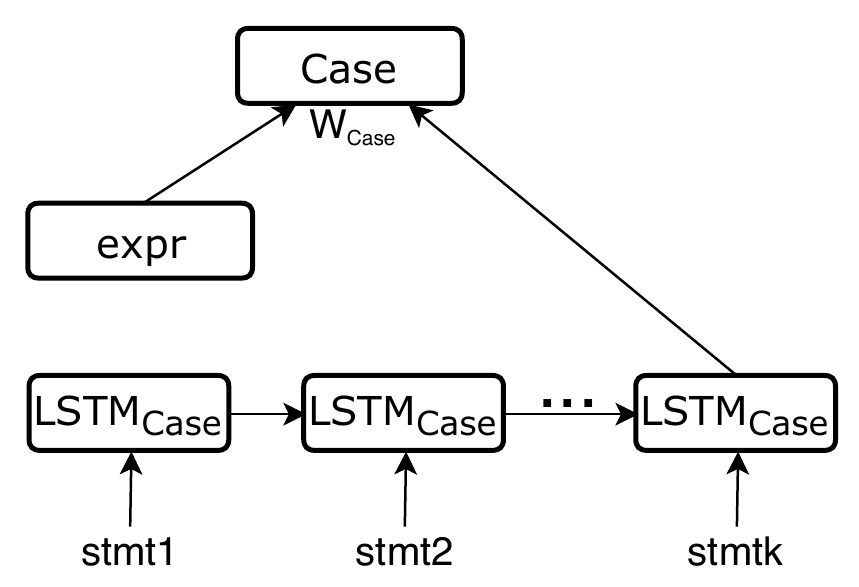} 
\vspace{-1em}
\caption{Structure of Case module.} 
\vspace{-1em}
\end{figure}
\subsubsection{Nodes with sequential children}
In C AST there are some other types of nodes whose children form a sequence of arbitrary length. A typical example is the Compound node, which denotes a code block, its children are a sequence of statements. We also use an LSTM to encode these children and take the last hidden state of the LSTM as the output.

\begin{table*}[!htb] 
\scriptsize
\renewcommand\arraystretch{1.2}
  \centering
    \caption{Neural modules for MTN. $LSTM()$ takes a sequence of input and output the last hidden state.}
    \begin{tabular}{|l|l|l|}
    \hline
        Semantic Unit  & Function & Parameters\\
    \hline
         {\fontfamily{\ttdefault}\selectfont FuncDef}$\rightarrow$ decl, body & $h_{FuncDef}=tanh(W_{FuncDef}[h_{decl}:h_{body}]+b_{FuncDef})$ & $2d^{2}+d$\\
    \hline
         {\fontfamily{\ttdefault}\selectfont While}$\rightarrow$ cond, stmt & $h_{While}=tanh(W_{While}[h_{cond}:h_{stmt}]+b_{While})$ & $2d^{2}+d$ \\
    \hline
         {\fontfamily{\ttdefault}\selectfont DoWhile}$\rightarrow$ cond, stmt & $h_{DoWhile}=tanh(W_{DoWhile}[h_{cond}:h_{stmt}]+b_{DoWhile})$ & $2d^{2}+d$ \\
    \hline
         {\fontfamily{\ttdefault}\selectfont For}$\rightarrow$ init, cond, next, stmt & $W_{For0}[tanh(W_{For1}[h_{init}:h_{cond}:h_{next}]+b_{For1}):h_{stmt}]+b_{For0}$ & $5d^{2}+2d$\\
    \hline
         {\fontfamily{\ttdefault}\selectfont If}$\rightarrow$ cond, iftrue & $h_{If}=tanh(W_{If}[h_{cond}:h_{iftrue}]+b_{If})$ & \multirow{2}{*}{$2d^{2}+d$}\\
    \cline{1-1} \cline{2-2} 
         {\fontfamily{\ttdefault}\selectfont If}$\rightarrow$ cond, iftrue, iffalse & $h_{If}=max(tanh(W_{If}[h_{cond}:h_{iftrue}]+b_{If}),tanh(W_{If}[h_{cond}:h_{iffalse}]+b_{If}))$ & \\
    \hline
         {\fontfamily{\ttdefault}\selectfont Switch}$\rightarrow$ cond, stmt & $h_{Switch}=tanh(W_{Switch}[h_{cond}:h_{stmt}]+b_{Switch})$ & $2d^{2}+d$\\
    \hline
         {\fontfamily{\ttdefault}\selectfont Case}$\rightarrow$ expr, stmt1, ..., stmtk & $h_{Case}=tanh(W_{Case}[h_{expr}:LSTM_{Case}(h_{stmt1},...,h_{stmtk})]+b_{Case})$ & $10d^{2}+5d$\\
    \hline
         {\fontfamily{\ttdefault}\selectfont Seq}$\rightarrow$ child1, ..., childk & $h_{seq}=LSTM_{seq}(h_{child1},...,h_{childk})$ & $8d^{2}+4d$\\
    \hline
    \end{tabular}%
    
  \label{tab:addlabel}%
\end{table*}%

In summary, for hidden size $d$, the number of parameters for all types of modules are $33d^{2}+16d$. So an MTN-a model contains $8d^{2}+4d+33d^{2}+16d=41d^{2}+20d$ parameters while an MTN-b model has $9(8d^{2}+4d)+33d^{2}+16d=105d^{2}+52d$ parameters.

\subsection{Calculate representation for code snippets by MTN}
In order to calculate vector representation of source code with MTN, we need to parse our code snippets into ASTs first. After we obtain the AST of a code snippet, we built an MTN for it with the same structure of its AST. In an AST of C, each semantic unit only corresponds to a single parent node type so that we can determine the type of a semantic unit only by the parent node in it. So for nodes in \{{\fontfamily{\ttdefault}\selectfont FuncDef}, {\fontfamily{\ttdefault}\selectfont While}, {\fontfamily{\ttdefault}\selectfont DoWhile}, {\fontfamily{\ttdefault}\selectfont For}, {\fontfamily{\ttdefault}\selectfont If}, {\fontfamily{\ttdefault}\selectfont Switch}, {\fontfamily{\ttdefault}\selectfont Case}\} we place corresponding MTN units in their position. For other nodes, we first check whether their children contains a sequence. If they do, we place $MTN_{seq}$ in the position of the parent of sequences. Else we place a traditional Child-Sum Tree-LSTM unit in their position. After we finish the building of MTN, we can calculate the hidden state of each node in a bottom-up way, and the hidden state of the root node is engaged as the representation of the whole program. The input vector $x$ at every node is their initial word embedding.

\section{Applications}
In the above section, we described the general process of calculating the vector representation of a C AST by MTN. In real application scenarios, we first use training data and ground truth labels to train MTN networks, then evaluate testing data on our trained mode. In this section we show the training and evaluation process of MTN on two program analysis tasks: program classification and code clone detection.

\subsection{Program Classification}
For the program classification task, we meant to classify a C code snippet by its functionality. Suppose we aim to classify source code into $M$ classes, after we get the code vector representation $v$, we calculate the $M$-dimension predicted result $\hat{y}$ by applying a fully-connected neural network layer:
\begin{displaymath}
  \hat{y}=Wv+b
\end{displaymath}
When training, we use the cross entropy loss function:
\begin{displaymath}
  \mathrm{loss}(\hat{y},y)=-\mathrm{log}(\frac{\mathrm{exp}(\hat{y}_{y})}{\Sigma_{j}\mathrm{exp}(\hat{y}_{j})})
\end{displaymath}
Where $y$ is the ground truth class label, an $M$-dimension one-hot vector. During evaluation, we predict the class label by picking the dimension with the biggest value from $\hat{y}$, i.e. \begin{math}\mathrm{argmax}_{k}\hat{y}_{k}\end{math}.

\subsection{Code Clone Detection}
The core task of code clone detection is to determine whether two code fragments are duplicate. In this case, these two code fragments are passed through two separate MTNs with shared parameters. After we get their representation $v_{1}$ and $v_{2}$, we measure their relatedness by compute their cosine similarity \begin{math}\hat{y}=\frac{v_{1}\cdot v_{2}}{|v_{1}|\cdot |v_{2}|}\end{math}. Because \begin{math}\hat{y}\in [-1,1]\end{math}, we set the ground truth of true clone pairs are set to 1, and -1 for false clone pairs. When training, the loss is computed by mean squared error:
\begin{displaymath}
  \frac{1}{d}\sum_{i=1}^{d}(y_{i}-\hat{y}_{i})^{2}
\end{displaymath}
Which is reduced to squared error since the dimension $d$ of $\hat{y}$ is 1. During the evaluation, we predict whether a pair of code fragments is a clone by:
\begin{equation}
isClone=\left\{
\begin{array}{rcl}
True      &      & {\hat{y}>0}\\
False     &      & {\hat{y}\leq 0}
\end{array} \right.
\end{equation}

\section{Experiments}
\subsection{Datasets}
\subsubsection{Program Classification}
Our dataset is collected from an online student programming website POJ\footnote{http://poj.org/}. A smaller version of this dataset is already utilized by multiple tasks \cite{wei2017supervised,wei2018positive,ben2018neural}. We consider programs to solve the same problem belong to the same class. Our dataset contains 58,600 C program files from 293 programming problems, each of which contains 200 files. On average each file contains 22.8 lines of code and 116.62 AST nodes when transferred into AST. We split these files by 8:1:1 as training, validation and test set. Table 2 shows the quantity of several kinds of nodes in the whole set. We find that the number of nodes with sequential children far more exceeds other structural nodes. {\fontfamily{\ttdefault}\selectfont For} and {\fontfamily{\ttdefault}\selectfont If} nodes have a very high frequency of appearances, while the counts of {\fontfamily{\ttdefault}\selectfont DoWhile} and {\fontfamily{\ttdefault}\selectfont Switch} nodes are rather small.

\begin{table}[htbp]
  \centering
    \caption{Statistics of our program classification dataset on the appearances of different structural AST nodes.}
    \begin{tabular}{|l|l|}
    \hline
    Node type & \multicolumn{1}{l|}{Count of appearances} \\
    \hline
    {\fontfamily{\ttdefault}\selectfont FuncDef}   & 67,806 \\
    \hline
    {\fontfamily{\ttdefault}\selectfont While}  & 13,315 \\
    \hline
    {\fontfamily{\ttdefault}\selectfont DoWhile} & 1,645 \\
    \hline
    {\fontfamily{\ttdefault}\selectfont For} & 102,870 \\
    \hline
    {\fontfamily{\ttdefault}\selectfont If} & 116,152 \\
    \hline
    {\fontfamily{\ttdefault}\selectfont Switch} & 625 \\
    \hline
    {\fontfamily{\ttdefault}\selectfont Seq} & 519,425 \\
    \hline
    \end{tabular}%
  
  \label{tab:addlabel}%
\end{table}%
\subsubsection{Code Clone Detection}
Code clones can be categorized into four types \cite{roy2007survey}: Type 1: Identical fragments except for variations in whitespace, comments, and layout. Type 2: Identical fragments except for variations in identifier names and literal values. Type 3: Syntactically similar fragments that differ at the statement level. The fragments have statements added, modified, or removed. Type 4: Syntactically dissimilar fragments that implement the same functionality. Type 4 clones are the hardest to detect, so we mainly focus on type 4 clones in our experiment.

In order to show the advantage of MTN for detecting type-4 clones, we engage the OJClone dataset \cite{wei2017supervised} for the clone detection task. OJClone contains 15 different programming problems, and each problem contains 500 source code files in C. The average lines of code (LoC) in OJClone is 35.25\cite{wei2017supervised}. We consider that two different code fragments from the same problem are a true type-3 or 4 clone pair, while two code fragments from different problems are a false clone pair. We split all 7,500 code fragments by 8:1:1 for training, validation, and testing to guarantee the code fragments in validation and test set do not overlap with those in the train set. As there are far more false clone pairs than true clone pairs, we apply up-sampling to the true clone pairs in the training set. For the training set, we randomly sample the same number of 20,000 positive clone pairs and negative pairs from the training code fragments. For the validation and test set, we each sample 5,000 pairs from their set of code fragments no matter they are positive or negative clone pairs.

\subsection{Experiment Settings}
We apply pycparser\footnote{https://github.com/eliben/pycparser} to generate ASTs for C programs. In pycparser, there are two types of nodes: identifier node and non-identifier node. Non-identifier nodes only contain a type, while identifier nodes contain a type and at least one (most times only one) value. In our dataset, the names and values of identifiers used by programmers are highly related to the problem description. Therefore, these details may provide extra benefits for classifying programs. Since we want to classify programs only by their algorithms, this information of identifiers needs to be pruned. So in our default settings, we only keep the node type as the input for both the program classification and the clone detection tasks. In order to show the effect of identifiers, we also add the results for MTNs when names or values are used as the input of identifier nodes.

In the program classification task, we set the dimension of node embeddings, LSTM hidden states of MTN-b model to 200. In the clone detection task, we set them to 100. The MTN-b model for program classification contains $105\times 200^{2}+52\times 200\approx 4,210K$ parameters. For tree-LSTM and MTN-a, we use larger hidden sizes so that their number of parameters are similar to MTN-b (e.g. to compensate the parameter differences, in the program classification task we set the hidden size of tree-LSTM to 720 and MTN-a to 320). We use adaptive moment estimation (ADAM) \cite{kinga2015method} optimizer with a learning rate of 0.001 to train our model. For all models, the node embeddings are initialized with random-value vectors and learned through the training process. We implement our model in PyTorch \footnote{https://pytorch.org/} and train the models on an NVIDIA Tesla P100 GPU.

For tree-structured neural network models like MTN and tree-LSTM, batch processing is difficult since we need to build different trees for different input data. However, we still want to optimize our model with mini-batches, so we perform manually gradient accumulation. In manual gradient accumulation, we first divide our dataset into batches, which is the same as conventional batch processing. For each input example in a batch, we build a network following the input AST structure, feed the input example to it and calculate the loss. Then we manually accumulate the gradient computed by each example and perform optimization on the accumulated gradient. In our experiment, we set the training batch size to 32. 
Currently, there exist some dynamic batching techniques which can batch dynamic neural networks by putting instances of the same operation together \cite{neubig2017fly}, but so far PyTorch does not support these techniques. In the future, if we require a higher running speed of MTN, we can reimplement our models with other deep learning libraries that support dynamic batching.

\subsection{Baselines}
In both tasks, we compare our MTNs to several sequential and tree-structured neural network models: 

\textbf{Sequential models}: we employ a standard LSTM network, a bi-directional LSTM, and a one-dimensional CNN as our baselines. For CNN, we follow the approach in \cite{kim2014convolutional}, which used multiple filter window sizes (3,4,5). In all these sequential models, we apply depth-first traverse to convert an AST into a sequence of tokens as input for the model, the same as \cite{bhoopchand2016learning,li2017code}.

\textbf{Tree-structured models}: we take child-sum Tree-LSTM \cite{tai2015improved} and TBCNN \cite{mou2016convolutional} as our baselines. We also compare our results to Code-RNN \cite{liang2018comment}, a modified recursive neural network which is originally used to generate comments for source code. Slightly different from \cite{liang2018comment}, we employ abstract syntax trees for input data instead of parse trees (concrete syntax trees). 

\textbf{Graph-based models}: we also compare our models with gated graph neural networks (GGNN)\cite{li2015gated}. GGNN has already been used in several programmming language-related tasks such as variable naming and identifying variable misuse \cite{allamanis2017learning,BrockschmidtAGP19}. In our experiment we basically follow the approach in \cite{BrockschmidtAGP19} to build program graphs on ASTs. As our approaches do not consider the identifier values, we omit the data flow edges and only use Child, NextToken and NextSib edges.

For the code clone detection tasks, we also include the baselines used by \cite{wei2017supervised}:

Deckard \cite{jiang2007deckard}: a popular AST-based clone detection tool based on a characterization of subtrees with numerical vectors.

DLC (Deep Learning for Code Clones): a deep learning-based approach proposed by White et al. \cite{white2016deep}. It converts ASTs into binary trees and uses recursive neural networks to represent them. Unlike \cite{white2016deep}, we train the model with supervised clone labels, same as MTNs.

SourcererCC \cite{sajnani2016sourcerercc}:  a token-based clone detector that achieved strong recall and precision on type 1-3 clones.

\subsection{Research Questions and Results}
We investigate these following research questions by showing and analyzing our experiment results:

\noindent{\bf RQ1: How does our model perform in the program classification task?}

Our results for the program classification task are shown in Table 3. Our models, MTN-a and MTN-b, both achieve results better than all sequential and tree-structured baselines. This indicate that in addition to tree structures, modelling differences among semantic units in ASTs can provide more insightful information to neural network models. After applying identifier information, the accuracy of MTN is further improved, which is consistent with our anticipation. An unexpected finding is that GGNN performs poorly on the program classification task, its accuracy even lower than sequential models. This is likely because the GGNN model cannot well handle the hierarchical relationship between AST nodes.

In the program classification task, we run tree-LSTM and each MTN model for 10 times and report their average accuracy. To determine whether the accuracies of MTNs are significantly higher than tree-LSTM, we run the Wilcoxon signed-rank test \cite{wilcoxon1964some} on tree-LSTM with both MTN-a and MTN-b. In both tests the p-value is smaller than 0.05, meaning that our improvement is significant. Although the average accuracy of MTN-a is slightly higher than MTN-b, our significance test shows that there is not a significant difference between their performances.

\begin{table}[htbp]
  \centering
    \caption{Classification accuracy for the 293-classes program classification task}
    \begin{tabular}{|lr|}
    \hline
    Method & \multicolumn{1}{l|}{Accuracy} \\
    \hline
    CNN   & 68.3\% \\
    LSTM  & 80.8\% \\
    Bi-LSTM & 80.7\% \\
    \hline
    Code-RNN & 64.8\% \\
    TBCNN & 79.0\% \\
    GGNN & 61.0\% \\
    Tree-LSTM & 85.2\% \\
    \textbf{MTN-a} & \textbf{86.5\%} \\
    MTN-b & 86.2\% \\
    \hline
    MTN-a w/ id & \textbf{92.6\%} \\
    MTN-b w/ id & \textbf{92.9\%} \\
    \hline
    \end{tabular}%
  
  \label{tab:addlabel}%
\end{table}%
Next, we compare the converging process among models on the program classification task. A model that converges faster can reach maximum accuracy in fewer training epochs, thus indirectly reduce its training time. We display the convergence of our model by recording the test accuracy after each epoch of training. Figure 8 illustrates the changing process of test accuracy during the training of MTN and Tree-LSTM. During the first epochs, MTN models show substantial improvements over Tree-LSTM. Compared to Tree-LSTM, MTN models can achieve an acceptable result in the early phase of training and reach maximum accuracy sooner.

\begin{figure}[htbp] 
\centering 
\includegraphics[height=5cm, width=9cm]{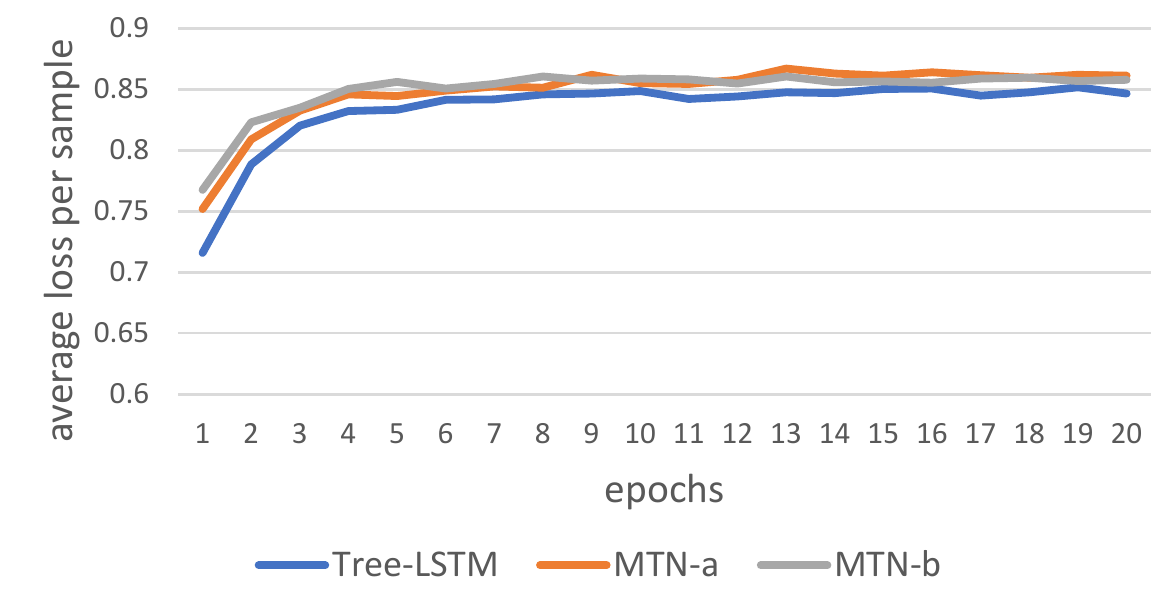} 
\caption{Test accuracy after every epoch of training in the program classification task.} 
\end{figure}

In real-world applications, it is often difficult to acquire a large number of programs with labeled categories to train neural network models, so we make further analysis on the performance of our models with fewer training data. We down-sample the training set by certain proportions and keep the validation and test set unchanged. The results of different down-sampling proportions are shown in Table 4. We can observe that when we shrink the training set, the performance gap between MTNs and Tree-LSTM becomes larger. This phenomenon indicates that MTNs are more advantageous for fewer training data.
\begin{table}[htbp]
  \centering
    \caption{Results of the program classification task when we sub-sample the training set by different proportions.}
    \begin{tabular}{cccc}
    \toprule
    \multicolumn{1}{r}{\multirow{2}{*}{Down-sampling proportion}} & \multicolumn{3}{p{12.57em}}{Results} \\
\cmidrule{2-4}          & \multicolumn{1}{p{6em}}{Tree-LSTM} & \multicolumn{1}{p{4.19em}}{MTN-a} & \multicolumn{1}{p{4.19em}}{MTN-b} \\
    \midrule
    12.5\% & 70.0\% & 72.6\% & 72.2\% \\
    \midrule
    25\%  & 77.1\% & 79.3\% & 78.4\% \\
    \midrule
    50\%  & 82.0\% & 83.2\% & 83.5\% \\
    \bottomrule
    \end{tabular}%
  
  \label{tab:addlabel}%
\end{table}%

\begin{table}[tbp]
  \centering
    \caption{Precision, recall and F1 score of the code clone detection task.}
    \begin{tabular}{|l|l|l|l|}
    \hline
    Method & \multicolumn{1}{l|}{Precision} & \multicolumn{1}{l|}{Recall} & \multicolumn{1}{l|}{F1} \\
    \hline
    Deckard & 0.6  & 0.06  & 0.11 \\
    \hline
    DLC   & 0.70  & 0.18     & 0.30 \\
    \hline
    SourcererCC & 0.97  & 0.1  & 0.18 \\
    \hline
    CDLH  & 0.21  & 0.97  & 0.34 \\
    \hline
    CNN   & 0.29  & 0.43  & 0.34 \\
    \hline
    LSTM  & 0.19  & 0.95  & 0.31 \\
    \hline
    Bi-LSTM & 0.18  & 0.97  & 0.32 \\
    \hline
    Code-RNN & 0.26  & 0.97  & 0.41 \\
    \hline
    GGNN & 0.20  & 0.98  & 0.33 \\
    \hline
    Tree-LSTM & 0.27  & \textbf{1.0} & 0.43 \\
    \hline
    MTN-a & 0.84 & 0.98  & 0.90 \\
    \hline
    MTN-b & \textbf{0.86}  & 0.98  & \textbf{0.91} \\
    \hline
    MTN-a w/ id & \textbf{0.91}  & \textbf{0.98}  & \textbf{0.95} \\
    \hline
    MTN-b w/ id & \textbf{0.91}  & \textbf{0.99}  & \textbf{0.95} \\
    \hline
    \end{tabular}%
  \label{tab:addlabel}%
\end{table}%

\noindent{\bf RQ2: How does our model perform in the clone detection task?}

Table 5 shows the result of the code clone detection task. For this task, we evaluate the precision, recall, and F1 score, where we set true clone pairs as positive samples. Similar to the program classification task, both MTN-a and MTN-b also outperform all the baselines in F1 scores. MTN models significantly outperform non-neural-network baselines in terms of recall, mainly because that MTN models are effective at detecting type-4 clones where the baselines failed. When compared to neural network models, MTNs achieve remarkable promotion in precision compared to most baselines. The performance of MTN-b is slightly higher than MTN-a. It is worth noting that most neural network baselines (e.g. Tree-LSTM) have unbalanced precision and recall. To find out the reason behind this, we take a look at the learned similarity scores between code pairs in Tree-LSTM and MTN. Figure 9 shows the distribution of similarity scores predicted by MTNs and tree-LSTM on the OJClone test set. We can see that for positive clone pairs, the prediction of tree-LSTM and MTNs are similar. However, for negative clone pairs, the behavior between tree-LSTM and MTN are highly different. While the predictions of MTN-a and MTN-b are close to -1, the predictions of tree-LSTM are much higher, which most of them fall in the interval of [-0.4,0.2]. Although we can improve the precision of tree-LSTM by increasing the threshold between positive and negative clones, the threshold must be chosen delicately since the gap of scores between positive and negative pairs are very small (in this case [0.4,0.6]). The selection of a data-specific threshold can be difficult if we do not have a proper validation set. On the other hand, the predictions of MTNs are close to 1 and -1, meaning that a simple choice of threshold (here 0) can be data-independent.

\begin{figure*}[htbp] 
\centering 
\includegraphics[height=7.5cm, width=15cm]{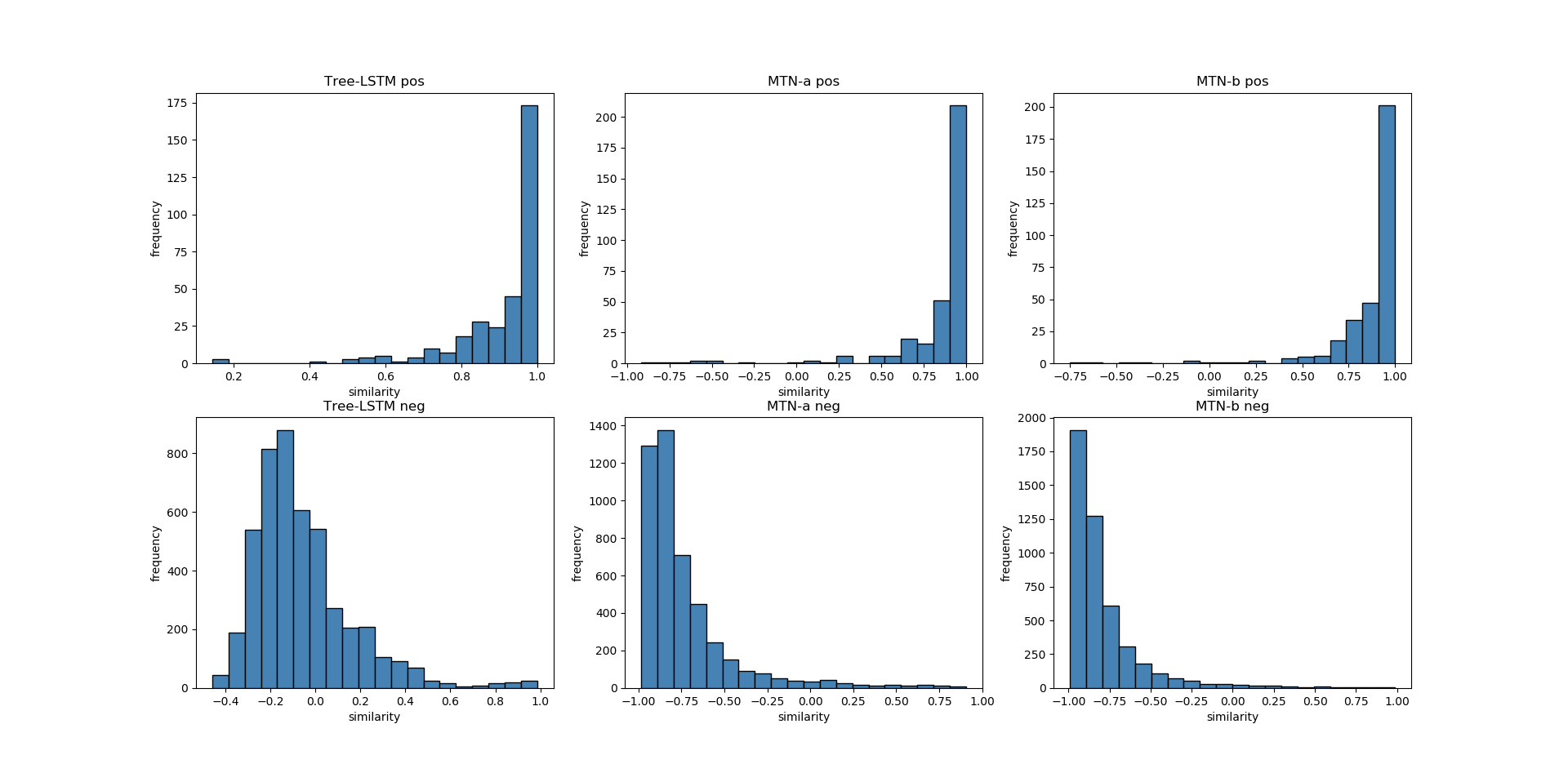} 
\vspace{-1em}
\caption{The distribution of similarity scores predicted by MTNs and tree-LSTM for the positive and negative clone pairs.}
\vspace{-1em}
\end{figure*}

\begin{figure*}[htbp] 
\centering 
\includegraphics[height=7.5cm, width=15cm]{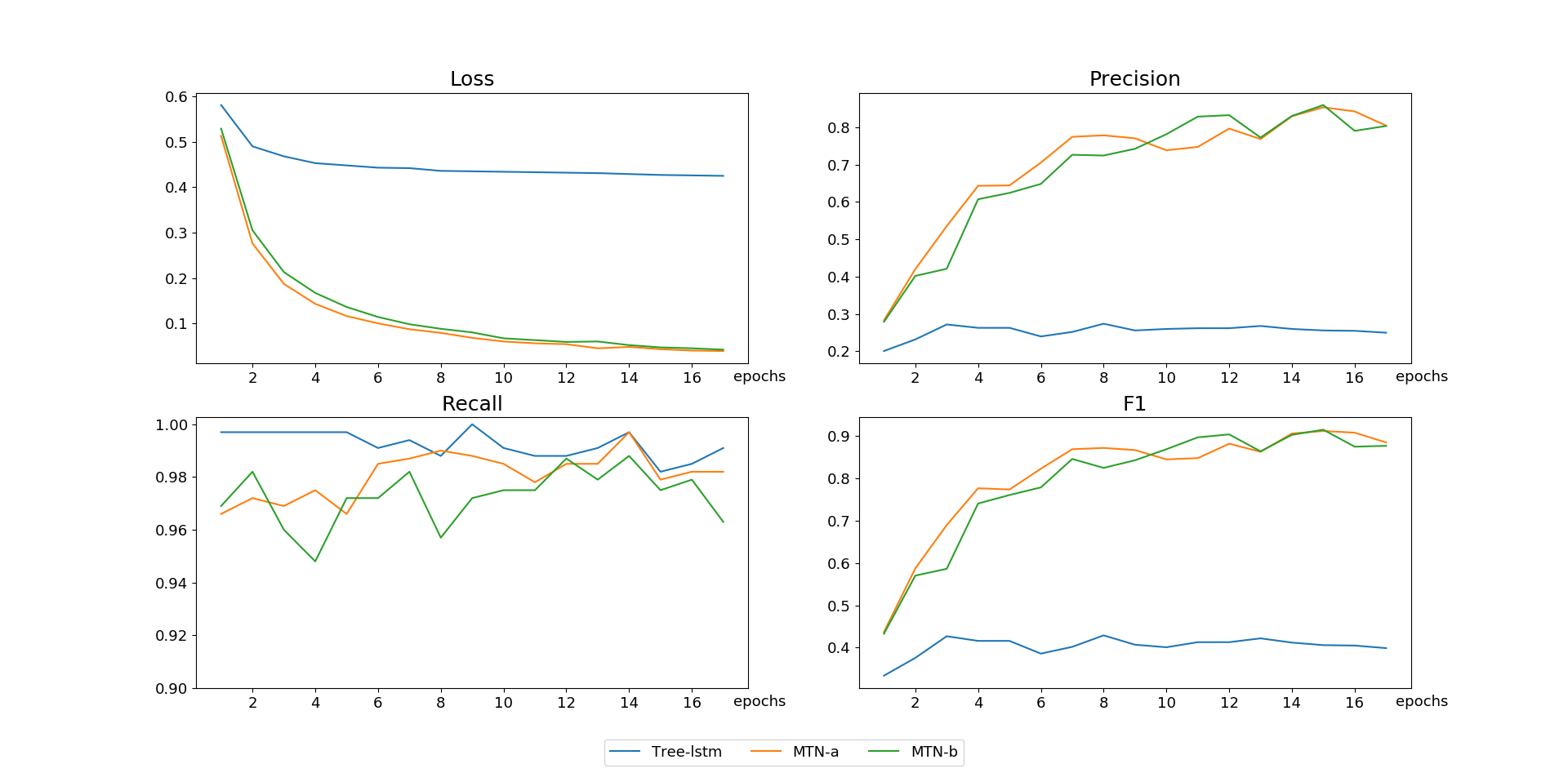} 
\vspace{-1em}
\caption{The training loss and the test set precision/recall/f1 score of MTN-a, MTN-b and tree-LSTM during the training procedure of the clone detection task.}
\vspace{-1em}
\end{figure*}

For the code clone detection task, we also analyze the convergence of MTN models. Figure 10 shows the average training loss per sample and the precision/recall/F1 score of the validation set during the training procedure. We can observe that compared to Tree-LSTM, MTNs can converge to a much lower loss. Although the loss of tree-LSTM is higher than MTNs, we make sure that tree-LSTM has come to convergence since, in our clone detection experiments, the loss of tree-LSTM finally drops to around 0.4 and does not drop further even if we continue to increase its hidden size (e.g., we run tree-LSTM on clone detection with hidden size 720, which is the same as the program classification task and the parameter size is nearly 4 times as the tree-LSTM for clone detection reported in this paper). Moreover, this higher loss than MTNs also exists in several other baselines, including both sequential models like LSTM and more powerful models like Code-RNN and GGNN, which further prove that this phenomenon is unlikely relevant to model capacity. The differences in learning speed are also indicated by the changing process of precision and f1. 

\begin{table}[htbp]
  \centering
    \caption{Detailed results for MTNs compared to Tree-LSTM on clone detection task.}
    \begin{tabular}{lcccc|c}
    \toprule
    Model & \multicolumn{1}{l}{p@r=0.8} & \multicolumn{1}{l}{p@r=0.9} & \multicolumn{1}{l}{p@r=0.95} & \multicolumn{1}{l}{p@r=0.99} &\multicolumn{1}{l}{ROC\_AUC}\\
    \midrule
    Tree-LSTM & 0.861 & 0.831 & 0.816 & \textbf{0.677} & 0.994 \\
    MTN-a & 0.984 & 0.970 & 0.934 & 0.470 & 0.995 \\
    MTN-b & \textbf{0.996} & \textbf{0.983} & \textbf{0.977} & 0.450 & \textbf{0.997} \\
    \bottomrule
    \end{tabular}%
  \label{tab:addlabel}%
\end{table}%

To have a more comprehensive understanding of the difference between our model and baselines on the clone detection tasks, we further compare MTNs with the best-performed baseline tree-LSTM on several other metrics, including the precision values for specific recalls, and area under ROC (Receiver-Operating Characteristic) curve. Table 6 shows the precision values for different recalls and the ROC\_AUC score for three models. The MTN-b achieves the highest ROC\_AUC among three models, and generally, its precision is higher than baseline models except for extreme circumstances when the recall is close to 1. We also draw the precision-recall curve for these three models in Figure 11. We can clearly see that when the precision is high (>0.8), the recall of MTN-a and MTN-b are significantly higher than tree-LSTM.

\begin{figure}[htbp] 
\centering 
\includegraphics[height=6cm, width=8cm]{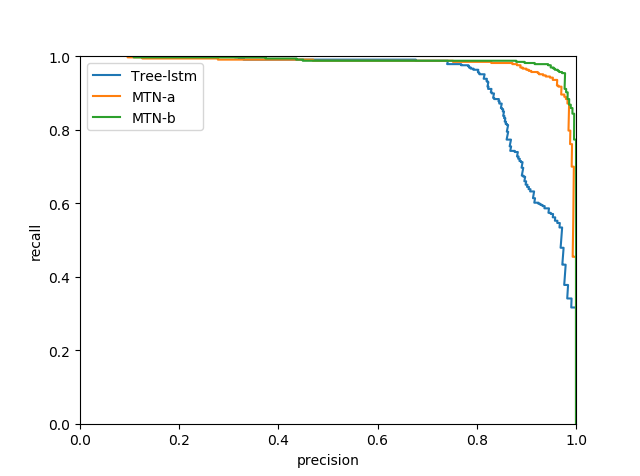} 
\vspace{-1em}
\caption{The precision-recall curve of MTNs and tree-LSTM on OJClone test set.}
\vspace{-1em}
\end{figure}

Additionally, we make a visualization of the embeddings of code snippets learned by MTN through clone detection. Figure 12 shows the distribution of 15 classes of programs in the OJClone dataset. We can find out that for programs from the same question, their embeddings are similar, and the boundaries between most questions are clear. So we can believe that MTN can learn semantic-aware embeddings for source code snippets even without direct class labels, which shows the potential of combining MTN with unsupervised learning.

\begin{figure}[htbp] 
\centering 
\includegraphics[height=7cm, width=9cm]{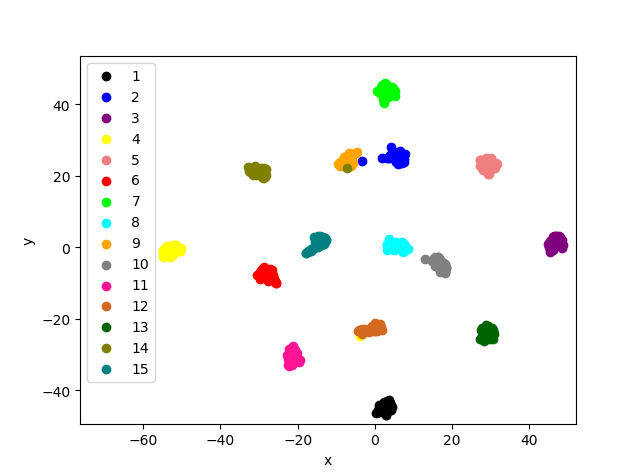} 
\caption{A 2D visualization of code embeddings learned by MTN-b in the clone detection task using t-SNE.} 
\end{figure}

Then, we will display some example clone pairs to show the capability of MTN intuitively:

Figure 13 shows 3 false clone pairs in OJClone that are wrongly predicted as positive by Tree-LSTM. We can see that when two code fragments have some similar patterns like they both include two function declarations (a), or both contains {\fontfamily{\ttdefault}\selectfont If} statements inside loops (b,c). From we previously mentioned, compared to MTNs, Tree-LSTM has a slightly higher recall and a much lower accuracy. So for code pairs with similarities Tree-LSTM have a higher chance to predict them as a clone, but MTNs can more accurately capture their semantics and give the correct prediction. 

Figure 14 shows a true clone pair which both Tree-LSTM and MTN-b predicted as false. Although these two programs solved the same question, their structure and algorithm are quite different. Even a human reader cannot quickly determine that they implemented the same functionality. 

Figure 15 shows a false clone pair which both Tree-LSTM and MTN-b predicted as true. Although the questions they aimed to solve are different, these two programs both involve structure and linklist operations. After a quick peeks, a human reader may find that they are very similar in structure.

\begin{figure*}[ht] 
\centering 
\includegraphics[height=13.5cm, width=13.5cm]{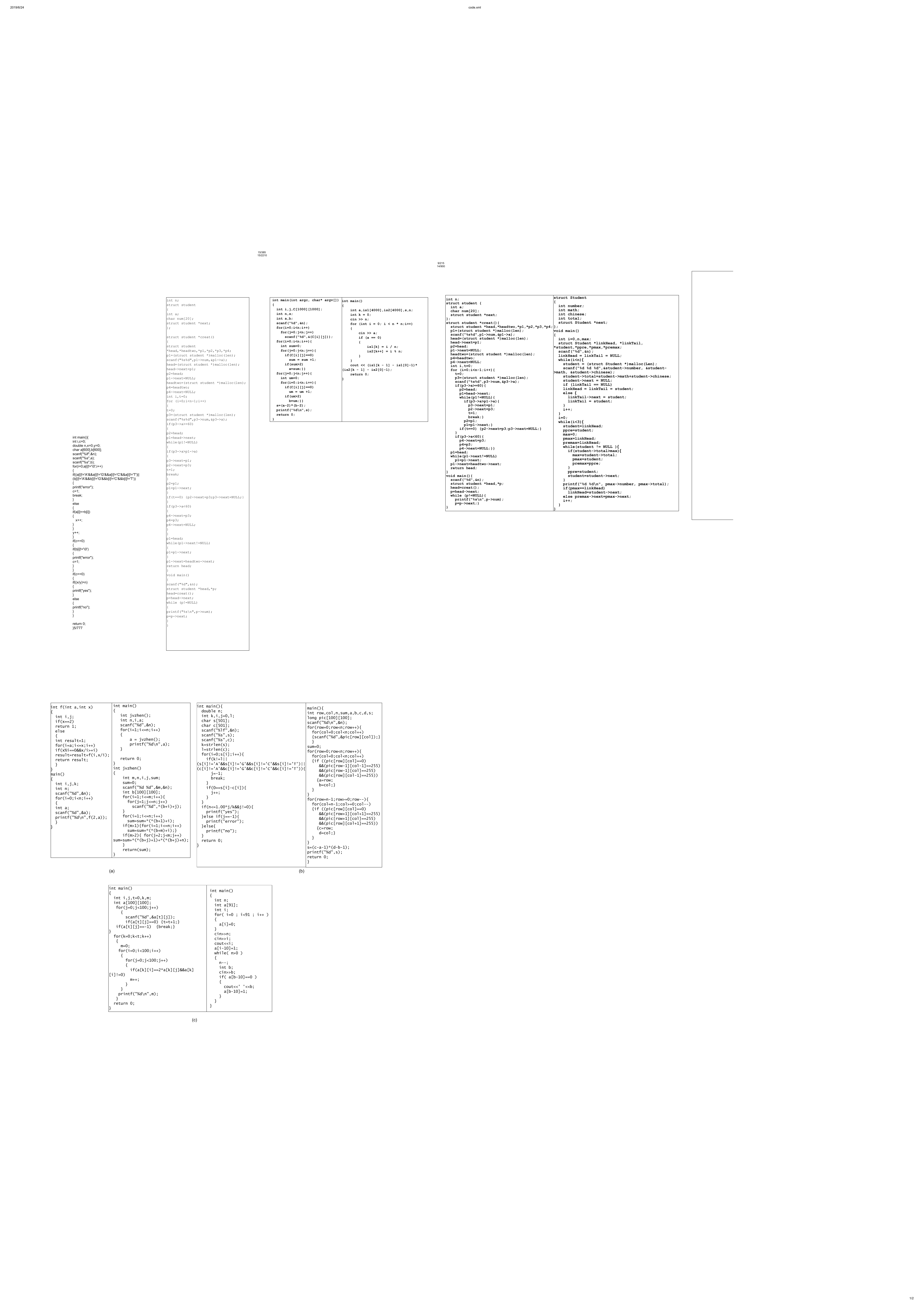} 
\vspace{-1em}
\caption{Examples of false clone pairs which MTN-b correctly predicted but Tree-LSTM failed.}
\vspace{-1em}
\end{figure*}

\begin{figure}[htbp] 
\centering 
\includegraphics[height=8cm, width=9cm]{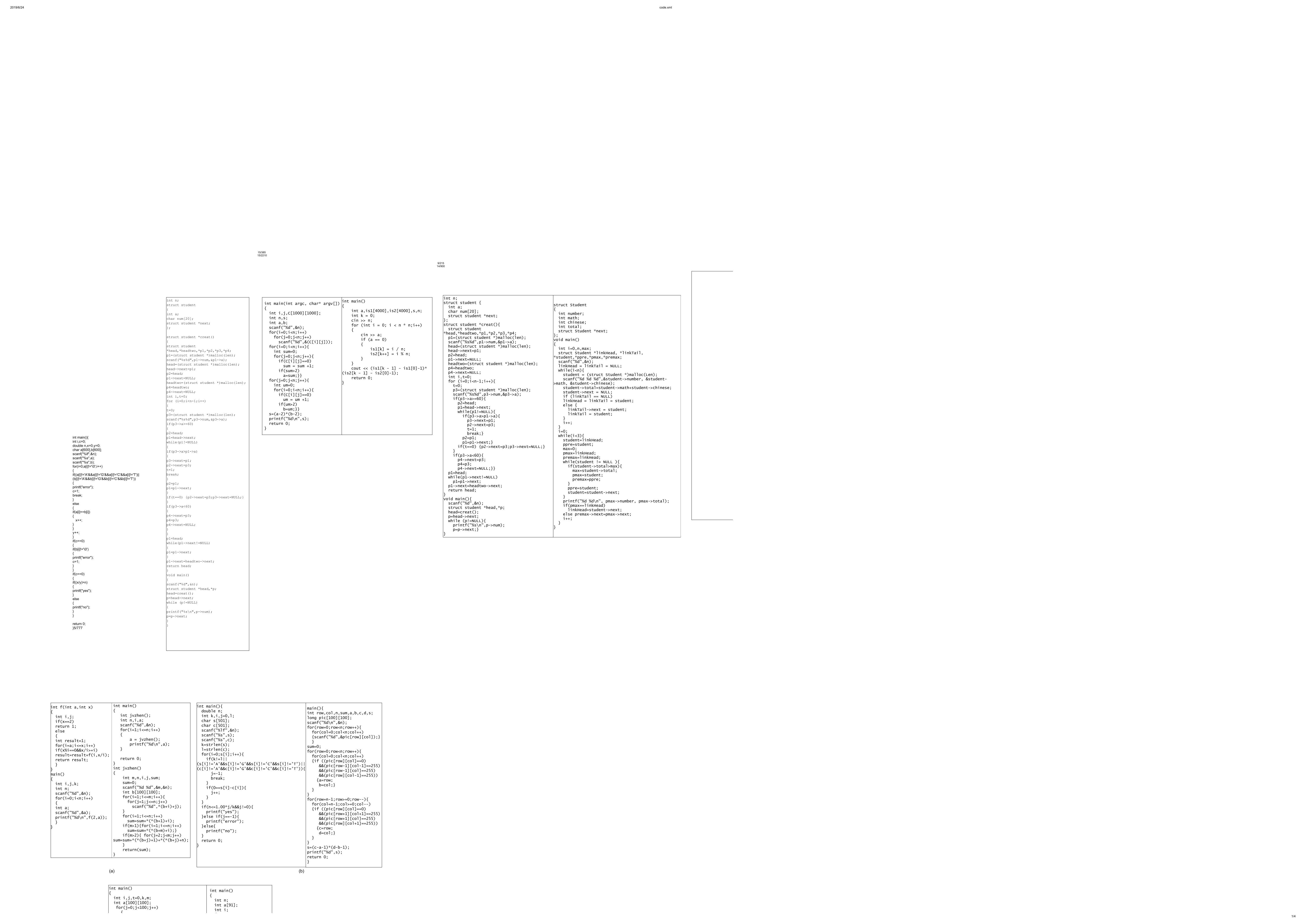}
\vspace{-1.5em}
\caption{An example of a true clone pair which both Tree-LSTM and MTN-b predicted as false.} 
\end{figure}

\begin{figure*}[htbp] 
\centering 
\includegraphics[height=12cm, width=12cm]{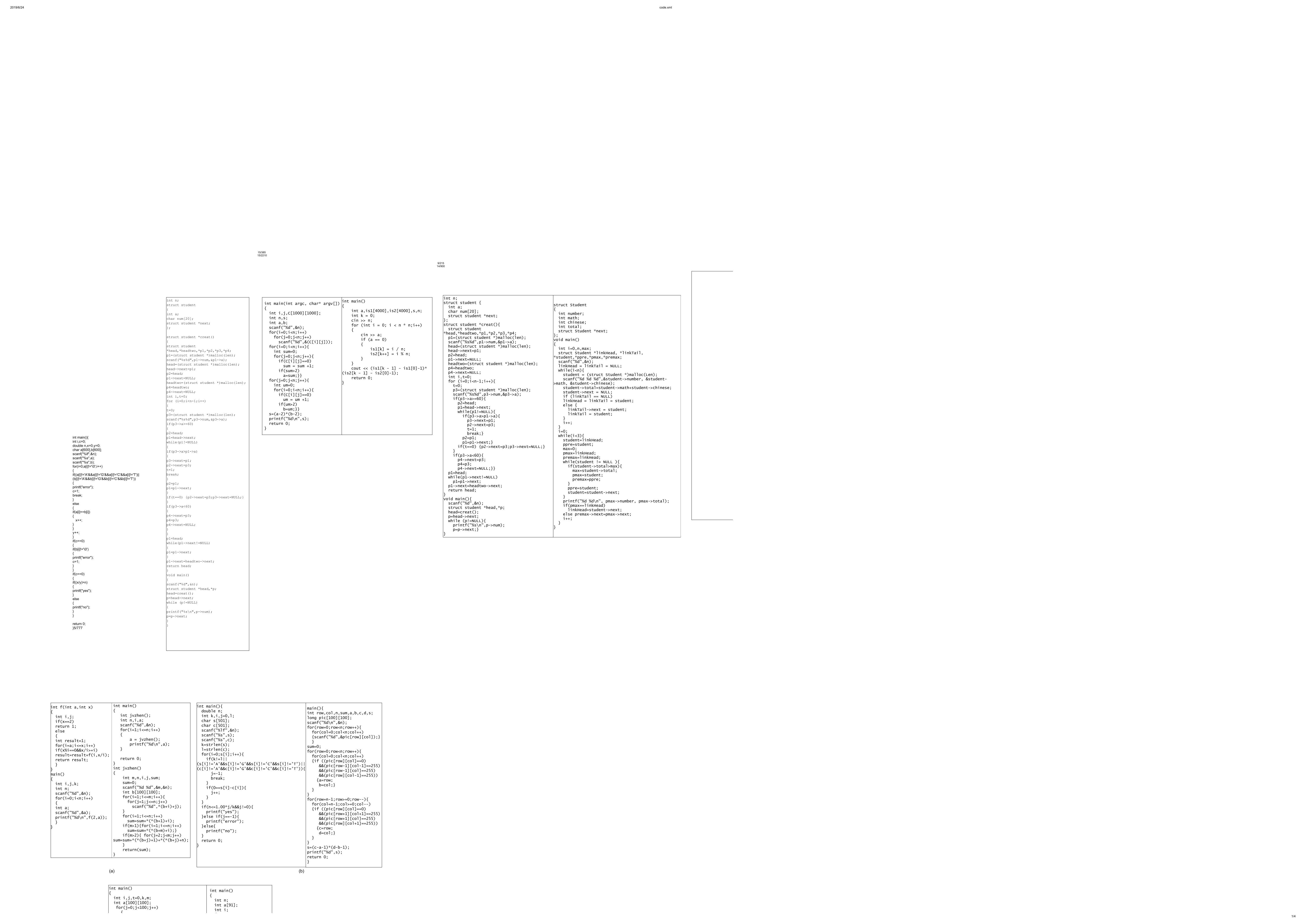}
\vspace{-1.5em}
\caption{An example of a false clone pair which both Tree-LSTM and MTN-b predicted as true.} 
\end{figure*}

\begin{table}[htbp]
  \centering
  \vspace{-1em}
    \caption{Ablation results for MTN-b on the clone detection task.}
    \begin{tabular}{lrrrc}
    \toprule
    Model & \multicolumn{1}{l}{precision} & \multicolumn{1}{l}{recall} & \multicolumn{1}{l}{f1}  & \multicolumn{1}{c}{ROC\_AUC}\\
    \midrule
    MTN-b & \textbf{0.859} & 0.975 & \textbf{0.913} & \textbf{0.997}\\
    -FuncDef & 0.826 & \textbf{0.991} & 0.901 & \textbf{0.997}\\
    -For  & 0.775 & 0.985 & 0.868 & \textbf{0.997}\\
    -If   & 0.779 & 0.985 & 0.870 & 0.996\\
    -While & 0.769 & 0.972 & 0.859 & 0.996\\
    -Seq  & 0.275 & \textbf{0.991} & 0.431 & 0.995\\
    \bottomrule
    \end{tabular}%
  \label{tab:addlabel}%
\end{table}%

\noindent \vspace{0.1cm} {\bf RQ3: How does each module of MTN contribute to our results?}

To analyze the contribution of each different MTN unit in our models, we perform an ablation study on the clone detection task. Each time we separately remove an MTN unit from our MTN-b model and replace it by a traditional child-sum tree-LSTM unit. We focus on five different MTN units which have higher frequencies than other units: $MTN_{FuncDef}$, $MTN_{For}$, $MTN_{If}$, $MTN_{While}$ and $MTN_{Seq}$. Table 6 shows the result of our ablation study. We can observe that removing $MTN_{Seq}$ has the biggest impact on the performance of MTN-b, and its precision and f1 only achieve slightly higher than tree-LSTM, implying that handling the order of statements and expressions are the key of MTN's improvements. For other MTN units, removing one of them only causes a slight drop in f1 or roc\_auc. This is probably because that the structures of subtrees corresponding to these units are fixed (e.g., an {\fontfamily{\ttdefault}\selectfont If} node always has a condition child and a loop body child), so a simpler model (tree-LSTM) can already learn enough information from these structures, and the effect of MTN units are limited.

\noindent \vspace{0.1cm} {\bf RQ4: What are the time and memory efficiency of MTN?}

We compare the time and memory efficiency of MTN models with tree-LSTM since they are all tree-structured models. In table 8, the first two rows show the running time of performing an epoch of training and testing on our clone detection task, and the third row shows the GPU memory cost during training. 
\begin{table}[htbp]
  \centering
    \caption{The time and GPU memory cost for MTN and tree-LSTM in our clone detection task.}
    \begin{tabular}{llll}
    \toprule
          & Tree-LSTM & MTN-a & MTN-b \\
    \midrule
    Training time (s) & 28871 & 29993 & 29366 \\
    Test time (s) & 1720 & 1810 & 1793 \\
    \midrule
    GPU memory cost (MB) & 743 & 731 & 721 \\
    \bottomrule
    \end{tabular}%
  \label{tab:addlabel}%
  
\end{table}%

We can see that when equipped with equal numbers of parameters, the training and testing time of MTNs are longer than Tree-LSTM only by a tiny scope (less than 6\%). Within two MTN models, the running speed of MTN-b is faster than MTN-a, which means that a larger hidden size of MTN-a brings more computational cost than MTN-b. When we look at the memory cost, the ranking for these three models is tree-LSTM > MTN-a > MTN-b. Since larger hidden size leads to larger intermediate tensors, this phenomenon fits our expectations. Generally, with the same number of parameters, MTN-b is more time and memory-efficient than MTN-a, and both models are nearly equally efficient to tree-LSTM.

\section{Threats to Validity}
\noindent {\bf Threats to internal validity} relates to the implementation of the CDLH baseline. In the paper of CDLH the authors did not give the specification of their model, so we reimplemented their model using PyTorch. The hyperparameters of CDLH is the same as our MTN models.

\noindent {\bf Threats to external validity} relates to the dataset we used and the generalizability of our model. Because the authors of \cite{wei2017supervised} did not give the specification of their OJClone dataset, we reproduced all their baselines and CDLH on our OJClone dataset. Still our approach improves them by a substantial margin. Since our approach mainly works for C-language, we evaluate it using OJClone. However, the idea of MTN is to build different neural modules for different grammar production rules, we can extend our model to OO-languages by building neural modules for OO production rules. In the future, we'd like to evaluate MTN on some OO-language benchmarks like BigCloneBench \cite{svajlenko2014towards}.

\section{Related Work}
Source code representation learning is a task that gets a distributed representation of a source code snippet through machine learning. In the field of source code representation learning, most models obtain distributed representations in a supervised manner. Supervised source code representation models are trained by tasks with labeled data, such as program classification, defect prediction, clone detection, and code summarization, and an intermediate result is obtained as the representation.

In this section, we first take a brief view of previous approaches in learning source code representations. In most of these tasks, the proposed models come from the natural language processing community. The neural network models for source code modeling can be divided into three categories: sequential, hierarchical and tree-structured models. Then we show some previous work on modeling different semantic compositions in non-programming language tree structures, which is slightly related to our work, and show their insufficiency for working on ASTs.

\subsection{Sequential Models}
Like natural language, source code snippets can be treated as sequences of tokens. Most investigations employ lexical analyzers to decompose source code into lexical tokens. Iyer et al. \cite{iyer2016summarizing} proposed Code-NN, which uses a Long Short-Term Memory (LSTM) \cite{hochreiter1997long} network with attention to generate natural language descriptions from C\# code snippets and SQL queries. Allamanis et al. \cite{allamanis2016convolutional} applied a multi-layer convolutional neural network with attention on Java source code to generate short, name-like summaries. Wang et al. \cite{wang2016automatically} engaged deep belief networks \cite{hinton2006fast} for defect prediction in several open source Java projects. Apart from models based on sequences of lexical tokens, \cite{bhoopchand2016learning,li2017code} serialized abstract syntax tree by depth-first traversal and applied recurrent neural networks to learn representations of code snippets for code completion. Although these sequential models were easy to implement and often fast to execute, they abandoned the explicit higher-level structural information of source code.

\subsection{Hierarchical Models}
Slightly more complicated than sequences, programs can also be represented as two-level hierarchies. Each program statement, often taking up one line of code, is a sequence of lexical tokens, and the whole program is a sequence of lines of code. Huo et al. \cite{huo2016learning} used one-dimensional convolutional neural networks for both within-line and cross-line modeling for program bug localization, while Huo et al. \cite{huo2017enhancing} applied CNN for within-line and LSTM for cross-line on the same task. Although these models utilized the sequential structure between lines of code, they failed to capture some kinds of logic structure within code lines, because lines of code in some program structures like loop and branches are not executed sequentially.

\subsection{Tree-structured Models}
Most tree-structured neural models for source code are built on programs ASTs. The basic component of these tree-structured models is a neural network unit which calculates the representation of a semantic unit. To represent tree structures, Socher et al. \cite{socher2011parsing} proposed recursive neural networks. In recursive neural networks, the vector embedding of a non-leaf node is calculated by its children with a fully-connected neural network, and the embedding of the root node is considered as the representation of the whole tree. However, traditional recursive neural networks require the trees to be binary, while the ASTs of source code are not. One way to apply recursive neural networks to ASTs is converting ASTs into binary trees \cite{white2016deep}, but the converting process make the ASTs much deeper and the model much harder to train. To adapt recursive neural networks for source code, Liang et al. \cite{liang2018comment} proposed Code-RNN, which operates on Java parse trees to learn source code representations for program classification and comment generation. Tai et al. \cite{tai2015improved} proposed Tree-LSTM, a tree-structured recursive neural network framework augmented with LSTM units. In a basic Tree-LSTM unit, hidden states of child nodes are passed into an LSTM-like NN unit and combined with the token embedding of the parent node to get the hidden states of the parent node. Tree-LSTM contains two types of model: Child-Sum Tree-LSTM and N-ary Tree-LSTM. Child-Sum Tree-LSTM can be built on trees with an indefinite branching factor, but it does not consider the order of children. N-ary Tree-LSTM is capable of capturing the order of children, but it requires that the branching factor of the tree must be known. Wei et al.  \cite{wei2017supervised} converted ASTs into binary trees, then applied binary Tree-LSTMs to detect code clones in Java and C. Mou et al. \cite{mou2016convolutional} introduced tree-based convolutional neural network (TBCNN), whose basic unit is a single dynamic convolution kernel over all semantic units of an AST. Unlike recursive networks, TBCNN did not calculate representations bottom-up, but by first calculating representations of semantic units separately, then applying pooling over all semantic units representations to get the final representation. They evaluated their model on tasks of program classification and bubble sort detection. These tree-structured networks all use the same neural unit with the same parameters for different semantic units. Apart from tree-structured models on ASTs, Allamanis et al. \cite{allamanis2017learning} augmented ASTs with edges of variable calls into a graph, then built gated graph neural network \cite{li2015gated} on which to solve variable misuse and variable naming tasks in C\#. Unlike other AST-based models, this model requires additional execution trace information which cannot be obtained by a static parser. Alon et al. \cite{alon2019code2vec,alon2018code2seq} extract syntactic paths from ASTs and use an attention mechanism to aggregate the embeddings of all paths.

\subsection{Modeling Different Semantic Compositions in Natural Language}
The differences among semantic units also exist in tree structures other than ASTs, like natural language parse trees. In these areas, researchers often use ``composition'' to refer to tree semantic units, mainly because in natural language constituency trees, a non-leaf node does not contain its own semantic information, where its semantics are composed by leaf nodes. Some researchers propose to use different compositional functions for different compositions \cite{socher2013parsing,dong2014adaptive,arabshahi2018combining}. For example, Socher et al.  \cite{socher2013parsing} proposed syntactically untied recursive neural network where each different sibling combination in a probabilistic contex-free grammar is assigned by a different weight matrix. Dong et al.  \cite{dong2014adaptive} Introduced AdaRNN, a recursive neural network with multiple compositional functions for sentiment classification. During the compositional phase, the composition result is calculated based on a probabilistic distribution over all composition functions. Arabshahi et al. \cite{arabshahi2018combining} modeled mathematic expressions with a Tree-LSTM in which each mathematical function was associated with a different set of parameters. In \cite{liu2017dynamic}, instead of defining separate composition functions in advance, it applied a meta-network of the same structure as the original network on the input data to produce dynamic parameters for the original network. Although these approaches pointed out that using different compositional functions for different semantic compositions is useful, the differences between compositional functions are only in the parameters. Moreover, most of these models only work for binary trees with simple structures, like natural language constituency trees, while program ASTs are often much deeper and wider. Thus these models cannot be properly applied to program ASTs.

\section{Conclusion and Future Work}
In this paper we propose MTN, a modular tree-structured recurrent neural network which can capture more detailed semantic information than previous tree-structured neural networks. We apply MTN in two tasks on C programming language: program functionality classification and code clone detection. Our results show that MTN outperformed previous sequential and tree-structured neural network models, due to its exploitation of semantic differences between AST semantic units. 

In the future, one of our priorities is to extend MTN to other programming languages. Another task worth investigating is to explore the potential of MTN for program generation. Most existing AST-based program generation tasks require the decoder to generate a program from the root node to leaves, so MTN cannot be used as the decoder for program generation. But MTN can be used to encode the partial tree during the generation process, provide the decoder with structural context information.


\begin{acks}
This research is supported by the National Key R\&D Program under
Grant No. 2018YFB1003904, the National Natural Science Foundation
of China under Grant No. 61832009, No. 61620106007 and No.
61751210, and the Australian Research Council’s Discovery Early
Career Researcher Award (DECRA) funding scheme (DE200100021).
\end{acks}

\bibliographystyle{ACM-Reference-Format}
\bibliography{sample-base}

\end{document}